\documentstyle[preprint,aps]{revtex}
\tightenlines

\begin{document}

\draft
\title
{Fractons and high-$T_{c}$ superconductivity}

\author
{Wellington da Cruz\footnote{E-mail: wdacruz@exatas.uel.br} 
and Marcelo Pagotto Carneiro}

\address
{Departamento de F\'{\i}sica,\\
 Universidade Estadual de Londrina, Caixa Postal 6001,\\
Cep 86051-970 Londrina, PR, Brazil\\}
 
\date{\today}

\maketitle

\begin{abstract}

We consider the concept of fractons in 
the context of high-$T_{c}$ superconductivity. These objects, which carry 
rational or irrational quantum numbers , are classified 
into universal classes $h$ of particles or quasiparticles which obey 
specific fractal distribution function. We show that 
the relaxation time associated to Hall conductivity for the 
superconducting cuprate systems came to out as 
$\tau_{H}\propto T^{-2}$. We also consider the pairing 
of fractons as a mechanism to produce bosonic systems 
and therefore superconducting states. For that an effective 
mass obtained from the propagator of a charge-flux 
system is considered. In this way, 
some experimental results of infrared studies of the cuprates 
for the effective mass, $m^*=m_{e}(1+\lambda)$, compared 
with our effective mass expression, $m_{eff}=m(1+s)$, 
show us that the dominant factor for interactions 
came from the spin. Thus spin flutuactions as a 
mechanism of high-$T_{c}$ superconductivity and 
fractons as quasiparticles are related. An expression to the 
low temperature specific heat of a quantum liquid of 
fractons is also obtained.

\end{abstract}

%\narrowtext

To date there is no consensus about the mechanism of high-$T_{c}$ 
superconductivity\cite{R1}. Among the models in the literature, the spin 
fluctuation models interest us as a possible mechanism in reason of the 
ideas exposed in this Letter\cite{R2}. Experimental 
results indicate that the 
transport properties of the cuprates in the presence or absence 
of magnetic fields 
have two separate scattering mechanisms with distinct 
relaxation times and so 
such models try to interpret the data.

In this paper, the concept of fractons$(h,\nu)$ is 
introduced and in continuing we show that a parallel 
discussion is possible between fractional quantum 
Hall effect-FQHE and superconductivity at high temperatures. 
Thus, some connection between fractons and the 
spin fluctuation models came to out.

{\it Fractons} are objects which carry rational or irrational 
quantum numbers, as charge and spin. These objects are classified 
into {\it universal classes $h$ of particles or quasiparticles}, 
where {\it $h$ is a fractal parameter or Hausdorff dimension}, 
associated with the path ( {\it fractal curve} ) of a 
quantum-mechanical particle. As a consequence, we have a 
{\it quantum-geometrical} description of the statistical 
laws of Nature. The particles within of the class $h$ obey 
specific {\it Fractal Statistics}\footnote{In fact, we have here 
{\it fractal functions}\cite{R3}.}( fractal distribution function ), 
given by\cite{R4}

\begin{eqnarray}
\label{e.45} 
n_{j}=\frac{1}{{\cal{Y}}[\xi]-h}\;,
\end{eqnarray}

\noindent where the function ${\cal{Y}}[\xi]$ satisfies 
the equation

\begin{eqnarray}
\label{e.46} 
\xi=\left\{{\cal{Y}}[\xi]-1\right\}^{h-1}
\left\{{\cal{Y}}[\xi]-2\right\}^{2-h}
\end{eqnarray}
 
 \noindent and $\xi=\exp\left\{(\epsilon_{j}-\mu)/KT\right\}$ 
has the usual definition. The parameter $h$ is defined into the interval
$1$$\;$$ < $$\;$$h$$\;$$ <$$\;$$ 2$ and the particles 
in each class $h$ are collected taked into account the {\it fractal 
spectrum}, which relates $h$ and the spin-statistics relation $\nu=2s$

\begin{eqnarray}
\label{e34}
&&h-1=1-\nu,\;\;\;\; 0 < \nu < 1;\;\;\;\;\;\;\;\;
 h-1=\nu-1,\;
\;\;\;\;\;\; 1 <\nu < 2;\;\\
&&h-1=3-\nu,\;\;\;\; 2 < \nu < 3;\;\;\;\;\;\;\;\;
h-1=\nu-3,\;\;\;\;\;\;\; 3 < \nu < 4;\;\nonumber\\
&&etc.\nonumber
\end{eqnarray}

\noindent We verify that the classes $h$ have a {\it duality
symmetry}, defined by

\begin{equation} 
\label{e.10}
\tilde{h}=3-h
\end{equation}

\noindent and so we can see that fermions$ 
(h=1)$ and bosons$(h=2)$ are dual objects. This suggests 
the concept of {\it fractal supersymmetry}, given that 
the supersymmetric pairs 
for energy states of spin $ ( s, s+\frac{1}{2} )$\cite{R5} have a realization 
through duality concept posed here.\footnote{ {\bf Theorem}: {\it If the 
particle with spin s is within the class h, then its dual 
$s+\frac{1}{2}$ is within the class ${\tilde{h}}$, so there exists 
a fractal supersymmetry}.}

Such ideas can be applied in the context of the FQHE.
This phenomenon is characterized  by the filling factor parameter $f$ 
, and for each value of $f$ we have the 
quatization of Hall resistance and a superconducting state 
along of the longitudinal direction of a planar system of electrons, 
which are manifested by semiconductor doped materials, i.e. heterojunctions, 
under intense perpendicular magnetic fields and lower 
temperatures\cite{R6}.

The parameter $f$ is defined by 
$f=N\frac{\phi_{0}}{\phi}$, where $N$ is the electron number, 
$\phi_{0}$ is the quantum unit of flux and
$\phi$ is the flux of the external magnetic field throughout the sample. 
The spin-statistics relation is given by 
$\nu=2s=2\frac{\phi\prime}{\phi_{0}}$, where  
$\phi\prime$  is the flux associated with the charge-flux 
system which defines the fracton$(h,\nu)$.

According to our approach there is a correspondence between
$f$ and $\nu$, numerically $f=\nu$. This way, 
 we verify that the filling factors observed 
 experimentally appear into the classes $h$ and from the 
 definition of duality
 between the equivalence classes, we note that the FQHE occurs in pairs 
 $(\nu,\tilde{\nu})$\cite{R4}.
 
 Going back to fractal distribution function expression, we can see that 
 for temperatures sufficientely low and 
  $\epsilon_{j}<\mu$, the average ocuppation number is given by
 $n=\frac{1}{2-h}$, and so the fractal parameter $h$ regulates
 the number of particles in each quantum state, i.e. 
 for $h=1$,$\;$$n=1$;$\;$$h=2$,$\;$$n=\infty$;$\;$$h=
 \frac{3}{2}$,$\;$$n=2$; etc. At $T=0$ and $\epsilon_{j}>\mu$, $n=0$
 if $\epsilon_{j}>\epsilon_{F}$ and $n=\frac{1}{2-h}$ 
 if $\epsilon_{j}<\epsilon_{F}$, therefore we have a {\it step  
 distribution}, taking into account the 
 Fermi energy $\epsilon_{F}$ and $h\neq 2$. 
 
 This behavior of the {\it fractal distribution function} 
 suggests us to consider a nearly ( {\it the 
 excitations are fractons} ) Landau-Fermi 
 liquid theory\cite{R7} in the 
 context of high-$T_{c}$ superconductivity to explain 
 the Hall conductivity time rate and so, we gain some 
 insight about the nature of the phenomenon in order. 
 On the other hand, we propose 
 {\it pairing of fractons} for the superconducting states. 
 In\cite{R8}, we considered 
 a continuous family of Lagrangians for a free charge-flux system 
 
 \begin{equation}
  \label{a14}
  L_{s}=\frac{{\dot x}^2}{2e}+\frac{s^2}{2e}
  +\frac{e}{2}(m^2-2m^2\alpha)
  \end{equation}
  
  \noindent and an expression for the propagator in the momentum space 
  was obtained

\begin{equation}
\label{a7}
{\tilde{\cal F}}(p)=\frac{1}{(p_{\mu}-mS_{\mu})^2-m^2}\;,
\end{equation}

\noindent where $s$ is an arbitrary parameter taked out of the 
Casimir of spin algebra, $ S^2=-s^2$ and  
$m$ is the mass of the charge-flux system. On one hand,  
we have that $ S_{\mu}\;p^{\mu}+\alpha m=0$, where $\alpha$ is the
helicity. From the propagator we can see 
that the charge-flux system, now subjected to an external magnetic 
field, has an {\it effective mass} 
given by ${m_{eff}}^2=m^2s^2+m^2-
2m^2\alpha^{\prime}$, such that 
$ S_{\mu}\;{p^{\prime}}^{\mu}+\alpha^{\prime} m=0$, 
with $ p^{\mu}\longrightarrow{p^{\prime}}^{\mu}=
p^{\mu}-\frac{e}{c}A^{\mu}$ and $\alpha^{\prime}$ 
embodied corrections which came from the interaction process. If we choose 
$\alpha^{\prime}=-s$ ( the precession angle of spin and momentum of 
the charge-flux system 
is the Hall angle between current and 
voltage in a magnetic field ), we obtain

\begin{eqnarray}
m_{eff}=m(1+s).
\end{eqnarray}

\noindent  Now we consider, this last expression for the system 
obtained by a pairing process of fractons. If we compare with the 
experimental effective mass, obtained from infrared 
studies of the cuprates, with effective mass formula\cite{R9}
 
\begin{eqnarray}
m^{*}=m_{e}(1+\lambda)\;,
\end{eqnarray}

\noindent where $\lambda$ is a parameter which taking into account 
many-body interactions, we note that for
$\frac{m^{*}}{m_{e}}=1,2,3,4$ our expression for the effective mass 
of the pairing process indicates that such superconductor 
systems are, in fact, bosonic 
systems with spin values $s=0,1,2,3$. Such matching between distinct 
expressions for the effective mass signalizes that the interactions 
are strongly dependent of the spin, hence the connection between 
the spin fluctuation models for high-$T_{c}$ superconductivity 
and our fracton model for quasiparticle. We observe here that $s$ can be 
any number, only in the superconducting state, pairing of fractons defines 
bosonic systems, otherwise we have fermionic and fractonic 
systems as alternative channels for the pairing process of fractons. 
In our analysis, the mechanism of high-$T_{c}$ 
superconductivity depends on spin in this route, 
given the parallel between one and the other effective mass formula.

The difference between the bosonic systems is related to degree 
of impurities of the samples. On the other hand, 
the step distribution for each class $h$ ( {\it remember that this one 
collects the particles or quasiparticles with distinct spin values 
which obey specific fractal distribution function} ), indicates that in the  
transition zone, around Fermi surface,  scattering of 
fractons (quasiparticles) has time rate 
$\tau_{H}\propto T^{-2}$, in accordance with experimental 
results\cite{R10}. This dependance of $\tau_{H}$ is verified at least in 
near optimal doping cuprates, where the Hall resistance 
is a strong function of temperature. We have that the collision probability 
of fractons is proportional to the square of the width of the 
transition zone, which is of the order of $KT$. Hence, the relaxation 
time is inversely proportional to the squared width. Also 
in\cite{R4}, at low temperatures $T$, we have obtained for the 
total energy

\begin{eqnarray}
{\cal{E}}&=&{\cal{E}}_{0}\left\{1+\frac{2h\pi^2}{6\mu_{0}^2}K^2T^2 \right\}\\
&=&{\cal{E}}_{0}\left\{1+\frac{\gamma V \pi^2}{6{\cal{E}}_{0}}K^2T^2\right\}\;, 
\end{eqnarray}

\noindent where $K$ is the Boltzmann's constant, $h$ is the Hausdorff 
dimension, $\mu_{0}$ and ${\cal{E}}_{0}$ are the zero-temperature 
chemical potential and energy respectively. We observe again that 
our analysis follow a type of Fermi liquid theory for near 
optimal doping cuprates ( in this sense see\cite{R11} ), 
where the elementary excitations 
are objects termed fractons. This is so because the 
experimental results suggest us such possibility. Thus, 
the low temperature specific 
heat of a quantum liquid of fractons 
is given by

\begin{eqnarray}
\frac{C[T,B]}{V}=\frac{\pi^2}{3}\gamma K^2 T \;,
\end{eqnarray}

\noindent with $\gamma=\frac{m_{eff}(\nu+1)}
{4\pi^2\hbar^2}$, $V$(area), $\nu=2s=\frac{\phi^{\prime}}
{2\pi\phi^{\prime}_{0}}$, $\phi^{\prime}_{0}=\frac{\hbar c}{2e}$ and 
$\phi^{\prime}=2\pi\nu\phi^{\prime}_{0}$ 
is a part of the external magnetic flux which is 
attached to the electron charge and so we define 
a fracton with $\nu$ a rational or irrational number ( for 
a recent discussion about fractional flux quantization in 
high-$T_{c}$ materials see\cite{R12} ). 
Therefore, the specific heat 
depends on the temperature $T$ and the magnetic field $B$, according to 
our model. 
There are in the literature another expressions to the low 
temperature specific heat of such superconducting systems\cite{R13}.

The longitudinal resistivity 
produces time rate $\tau_{\rho}$ which is proportional to $T^{-1}$, 
by kinetic theory. Another discussion about a parallel between FQHE and 
high-$T_{c}$ superconductivity was given in\cite{R14}.

To summarize, we have proposed that along of the longitudinal 
direction of the samples {\it pairing of fractons} produces bosonic 
systems and so superconducting states. Along of the transverse direction 
fractons define Hall states. Such simultaneous events, 
experimentally observed in the FQHE, can be considered in the 
context of high-$T_{c}$ superconductivity for justify 
the two time rates for the scattering mechanisms. The qualitative 
and in some sense quantitative  analysis considered here were 
based on the {\it fractal distribution function}( Fractal Statistics ) and 
an {\it effective mass} obtained from the propagator 
of a charge-flux system which defines the physical 
system. Our effective mass 
was compared with another one of infrared studies 
of the cuprate superconductors and an expression for 
the low temperature specific heat of such systems also was derived.

Finally, the FQHE and 
the high-$T_{c}$ superconductivity\cite{R14}, 
in spite of being distinct phenomena, 
can be analysed from the viewpoint of the fractons and 
are seen to overlap at some points. At another level, 
we would like to observe that the spin-charge separation 
ideas ( see Anderson in\cite{R1} ) do not are 
discharged of our formulation ( {\it pairing of fractons 
can go to fermions, bosons or fractons }) and so we may speculate about some
{\it hybrid theory} ahead to explain all the experimental facts.

\end{document}